\documentclass[aps,prl,twocolumn,superscriptaddress]{revtex4}
\usepackage{amsmath}
\usepackage{amssymb}
\usepackage{graphicx}

\begin{document}

\title{Kondo effect and RKKY interaction in magnetic trimers}

\author{Philipp Knake}
\author{A. L. Chudnovskiy}
\affiliation{1. Institut f\"ur Theoretische Physik, Universit\"at Hamburg,
Jungiusstr 9, D-20355 Hamburg, Germany}

\date{\today}

\begin{abstract} 
Clusters of magnetic atoms lying on a metallic substrate can exhibit Kondo effect due to the exchange interaction with mobile electrons in the metal, which can be observed in STM measurements. The same interaction results in the induced RKKY interaction between the magnetic moments in the cluster. For different geometries of the cluster, it can enhance or suppress Kondo-correlations. We calculate the tunneling density of states for trimer clusters of different geometries and reveal the signatures of competition between the Kondo effect and RKKY interaction that can be seen in STM experiments. 
\end{abstract}


\maketitle

Competition of Kondo correlations and RKKY interactions has a profound effect on  physics of magnetic alloys and artificially created structures such as quantum dots or  clusters of magnetic atoms \cite{genericKondo,Tsvelick,trimer,DQD}. 

Despite their common origin in the exchange interaction between localized magnetic impurities and mobile electrons, the Kondo correlations and RKKY interaction lead to quite opposite behavior at low energy scales. Kondo correlations lead to  the screening of localized magnetic moments by the spins of mobile electrons, resulting in the formation of a so-called Kondo cloud around each magnetic impurity. The Kondo screening is accompanied by a drastic rearrangement in the energy spectrum of mobile electrons: the Kondo peak in the one particle density of states (DoS) appears at the Fermi energy. This in turn leads to enhanced scattering of electrons by the screened magnetic impurity, which is registered experimentally as increase of the resistance of a metal with lowering temperature, or as increase of conductance through the artificial quantum dot \cite{genericKondo,QDKondo}. 
In contrast RKKY interaction leads to magnetic correlations between distant magnetic impurities, thus prohibiting their screening. While changing its sign with the distance, RKKY interaction can favor ferro- or antiferromagnetic ordering, or result in the spin glass order in alloys with random positions of magnetic impurities \cite{aboutRKKY}. In the case of dominating antiferromagnetic RKKY interaction, the Kondo effect is suppressed. 

Theoretical investigations of the competition between the Kondo effect and RKKY interaction mostly focused on the model of two magnetic impurities in a metal \cite{Wilkins}. Inspired by experiments on double quantum dots in Kondo regime 
\cite{DQD} This model was further extended in application to the transport through double quantum dot \cite{Vavilov}. There is a general result of theoretical considerations that strong RKKY interaction binds the two magnetic moments into a singlet state and thus suppresses the Kondo effect.  
 
The interplay between Kondo effect and RKKY interaction becomes even more involved in frustrated systems, such as clusters of three magnetic atoms with antiferromagnetic RKKY interaction. Artificial fabrication of such clusters deposited on a metallic substrate has become possible using the scanning tunneling microscopy (STM) \cite{trimer}. Moreover, STM technique allowed a controllable change of the positions of atoms and hence controlled change of the geometry of the trimer. 
The experiments showed clear dependence of low-energy density of states in the trimer on its geometry, which can be attributed to the competition of the Kondo effect and RKKY interactions. Subsequent theoretical investigations by means of continuous time Monte Carlo methods also showed the change of the density of states with the geometry of a triangular atomic cluster, and in particular the suppression of the Kondo peak in DoS for the antiferromagnetic sign of RKKY coupling \cite{Savkin}.

\begin{figure} 
\includegraphics[width=8cm,height=7cm,angle=0]{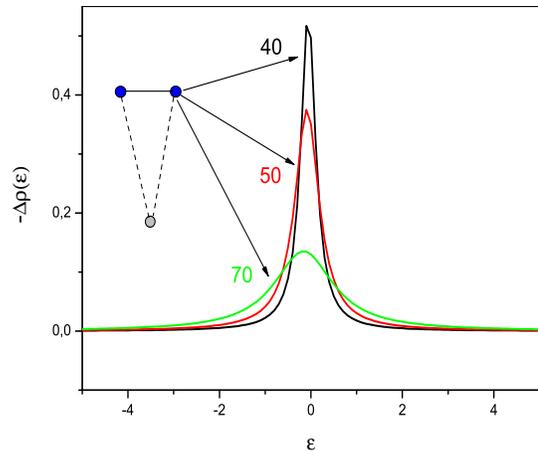}%
\vskip -1cm
\caption{Color online. Induced density of states for a strongly asymmetric isosceles triangle. The asymmetry of the triangle is fixed by the ration $g_2/g_1=40, 50, 70$ in the order of the diminishing height of the peak. The parameter $g_1=0.01 g_0$ remained fixed. 
\label{fig-dimer}}
\end{figure}

\begin{figure} 
\hskip -1cm\includegraphics[width=9cm,height=6.5cm,angle=0]{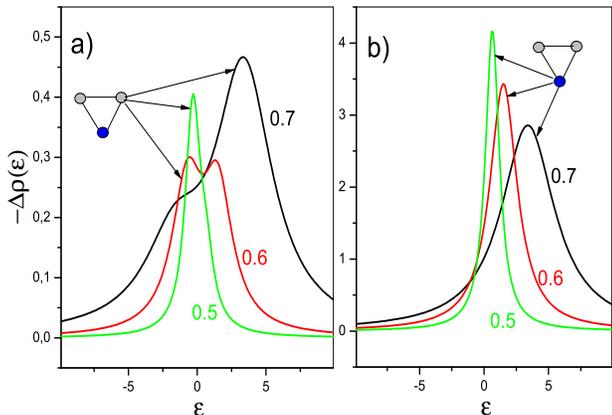}%
\vskip -1cm
\caption{Color online. Induced density of states for almost equilateral triangle ($G_1=1.01 g_1$) for $g_1/g_0=0.5, 0.6, 0.7$. The panel a) shows DoS on one of the atome forming a dimer. The panel b) shows DoS on a singled out atom.    
\label{fig-trimer}}
\end{figure}

Because of the common origin of the Kondo correlations and RKKY interaction it is important to treat both factors on equal footing. In this Letter we present such a treatment based on the mean field theory approach similar to the one used for the Kondo effect in the two level quantum dot \cite{Eto}. 
We present results on the local density of states (DoS) at positions of magnetic impurities, concentrating on the antiferromagnetic (AFM)  RKKY interaction. There are two qualitatively different regimes that are discriminated by the relative strengths of Kondo effect and RKKY coupling. In the Kondo regime, which is realized by weak RKKY interaction, there are pronounced Kondo peaks in DoS at positions of the magnetic atoms. The width and the hight of the peaks is modified by RKKY interactions and reflects the geometric form of the cluster. For an isosceles triangle we find a general tendency of the separation of the cluster into a single atom and the dimer with growing  RKKY interaction. The behavior at strong AFM RKKY coupling depends very much on the geometry of the cluster. In a strongly asymmetric cluster,  two atoms form a singlet state with strongly suppressed Kondo effect, while the distant atom is singled out and exhibits a Kondo peak in the local DoS. 
In contrast, an almost equilateral triangle shows a complex evolution of DoS with increasing RKKY interaction. At two atoms that are closer to each other,  the Kondo peak first diminishes and splits, and then grows with RKKY coupling. At the same time there is a growth in the local DoS at the third atom. Moreover the positions of the peaks in DoS deviates from the Fermi level.   Our findings are summarized in Figs. \ref{fig-dimer},  \ref{fig-trimer}. 

We consider a cluster of three magnetic atoms each having a spin 1/2 placed on a metallic substrate. The microscopic description of the system is given by the Hamiltonian of s-d model generalized to the case of three magnetic impurities 
\begin{eqnarray}
\nonumber 
&& \hat{H} =\sum_{{\bf k}\sigma}\epsilon_{\bf k} \hat{a}^{\dag}_{{\bf k}\sigma}\hat{a}_{{\bf k}\sigma}+\sum_{j\sigma}\epsilon_d \hat{c}^\dag_{j\sigma}\hat{c}_{j\sigma}\\ 
&&+\sum_j U \hat{n}_{j\uparrow} \hat{n}_{j \downarrow}+\sum_j\sum_{{\bf k}\sigma}\left(T_{j{\bf k}}\hat{c}_{j\sigma}^\dag \hat{a}_{{\bf k}\sigma}+\,\textrm{h.c.}\,\right)
\label{s-d-model}
\end{eqnarray}
Here operators $\hat{c}_{j\sigma}, \hat{c}^\dag_{j\sigma}$ relate to the localized d-level in the atom $j$, the operators $\hat{a}_{{\bf k}\sigma}, \hat{a}^{\dag}_{{\bf k}\sigma}$ relate to the mobile fermions in the metallic substrate. The hybridization between the localized levels and the extended states in the substrate is given by the last term in (\ref{s-d-model}). 
The RKKY interaction results from the finite overlap of the extended state ${\bf k}$ with the localized states of different atoms, that is $T_{j{\bf k}}\neq 0$ and $T_{j'{\bf k}}\neq 0$ for $j\neq j'$. To take into account RKKY interaction on equal footing with the Kondo effect, we formally extend the basis of states for mobile fermions introducing a three-component field for each extended state 
$\vert {\bf k} \sigma\rangle $ with components 
\begin{equation}
\hat{\psi}_{j{\bf k}\sigma}=\hat{a}_{{\bf k}\sigma} t_{j{\bf k}}, 
\label{psi} 
\end{equation}   
where $t_{j{\bf k}}=T_{j{\bf k}}/T_0$ with 
$T_0({\bf k})=\sqrt{\sum_i \vert T_{j{\bf k}}\vert^2}$.   Note that this formal extension does not introduce any additional degrees of freedom. 
The free Green's function of the fields $\psi_{j{\bf k}\sigma}$ in Matsubara frequency space becomes a matrix with elements 
\begin{equation}
\left[{\bf G}(i\omega_n, {\bf k})\right]_{j\sigma, j'\sigma'}=\frac{1}{i\omega_n-\epsilon_{\bf k}} t^*_{j'{\bf k}} t_{j{\bf k}} \delta_{\sigma\sigma'}.  
\label{G-omega}
\end{equation}
The coupling of the fields $\hat{\psi}_j$ and $\hat{\psi}_{j'}$ takes into account the elementary processes responsible for the RKKY interaction.  
After performing a Schrieffer-Wolff transformation, we obtain the effective Kondo Hamiltonian, which can be represented in the form 
\begin{equation}
H= \sum_{{\bf k}\sigma}\epsilon_{\bf k} \hat{a}^{\dag}_{{\bf k}\sigma}\hat{a}_{{\bf k}\sigma}+  J\sum_{{\bf k}, {\bf k'}}\sum_j\left[\Psi^{\dagger}_{j\bf k}{\bf \sigma}
\Psi_{j{\bf k'}}\right]\cdot \left[\Phi_j^{\dagger}{\bf \sigma}
\Phi_j\right], 
\label{HKondo}
\end{equation}
where $\Psi_{j{\bf k}}=(\hat{\psi}_{j{\bf k}\uparrow}, \hat{\psi}_{j{\bf k}\downarrow})^T$, $\Phi_j=(\hat{c}_{j\uparrow}, \hat{c}_{j\downarrow})^T$, and  the Kondo coupling is given by the standard expression \cite{Hewson}
\begin{equation}
J=T_0^2\left(\frac{1}{\epsilon_d+U}-\frac{1}{\epsilon_d}\right).  
\label{JKondo}
\end{equation}
In deriving (\ref{JKondo})  we assumed the average tunneling $T_0({\bf k})$ to be independent of the wave vector.    
To develop a mean field treatment of the problem, we write down the partition function for the Hamiltonian (\ref{HKondo}), and decouple the interaction term with the help of a Hubbard-Stratonovich transformation introducing the hybridization fields as follows 
\begin{eqnarray}
\nonumber &&
\exp\left\{J\sum_{\nu}\sum_{{\bf k}{\bf k'}}\left[\bar{\Phi}_j\sigma^{\nu}
\Phi_j\right]\left[\bar{\Psi}_j\sigma^{\nu}
\Psi_j\right]\right\}= \\
\nonumber &&
\int db_j \exp\left[-\frac{6}{J}\left(b_j\right)^2 
+3b_j\sum_{{\bf k}{\bf k'}}
\left(\bar{\Psi}_{j{\bf k}}\cdot\Phi_{j}+\bar{\Phi}_{j}\cdot\Psi_{j{\bf k'}}
\right)
\right]. \\
\label{HS1}
\end{eqnarray}
The Schrieffer-Wolff transformation reduced the Hilbert space at each magnetic impurity to the two spinful states. This allowed us to use  a semifermionic representation of spin degrees of freedom at each impurity in the path integral formulation of partition function \cite{Popov-Fedotov,Kiselev}. In the semifermionic representation, the propagator of the corresponding Grassman fields $\Phi_j$ is given by 
\begin{equation}
D_{j\sigma,j'\sigma'}(i\omega_n)=\frac{1}{i\omega_n+i\pi T/2}
\delta_{jj'}\delta_{\sigma\sigma'}, 
\label{D}
\end{equation}
where $T$ denotes the temperature. 
We made a static mean-field ansatz for the decoupling fields, $b_j^{\tau}=b_j$ ($\tau$ denotes the imaginary time). Then, after integrating out the fermionic variables, the partition function acquires the form $Z=e^{-F/T}$ with the free energy $F$ given by 
\begin{equation}
F=F_0+\frac{6}{J}\sum_{j=1}^3 b_j^2-3 T\sum_{\omega_n}{\mathrm{Tr}} 
\ln\left\{{\bf 1}_6+ {\bf D B \sum_{\bf k}G_k B} \right\}. 
\label{F} 
\end{equation}
Here $F_0$ is an additive constant independent of decoupling fields $b_j$, 
and the matrices are given by the following expressions: 
${\bf D}(\omega_n)=(i\omega_n+i\pi T/2)^{-1} {\bf 1}_6$, 
${\bf B}={\mathrm{diag}}(b_j)\otimes {\bf 1}_2$. The elements of the matrix 
$\sum_{\bf k}{\bf G_k}(i\omega_n)$  read 
\begin{equation}
\sum_{\bf k} G_{j\sigma,j'\sigma'}(i\omega_n, {\bf k})=
\sum_{\bf k}\frac{t^*_{j'{\bf k}} t_{j{\bf k}}}{i\omega_n-\epsilon_{\bf k}} 
\delta_{\sigma\sigma'}\equiv \tilde{g}_{jj'}(i\omega_n).  
\label{g}
\end{equation}
The matrix elements $\tilde{g}_{jj'}$ depend on the geometry of the magnetic cluster. Taking the positions of atoms as ${\bf r}_j$, we can write 
$t_{j{\bf k}}=e^{-i {\bf k r}_j}/\sqrt{3}$. The explicit expression for $\tilde{g}_{jj'}(i\omega_n)$ now reads 
\begin{equation}
\tilde{g}_{jj'}(i\omega_n)=\int\frac{d^3 {\bf k}}{3 (2\pi)^3} \frac{e^{i{\bf k}( 
{\bf r}_{j'}-{\bf r}_{j})}}{i\omega_n-\epsilon_{\bf k}}
=\int_0^{\infty}\frac{k dk}{6\pi^2 r_{jj'}} \frac{\sin(k r_{jj'})}{ i\omega_n -\epsilon_k}. 
\label{g_jj'}
\end{equation}  
Here $r_{jj'}=\vert {\bf r}_j-{\bf r}_{j'}\vert$, and we took into account that the energy $\epsilon_{\bf k}$ depends only on the absolute value $k$. The actual value of $\tilde{g}_{jj'}$ depends on the coordinates of magnetic atoms in an oscillatory manner and carries the information about RKKY equation.  Approximate evaluation of (\ref{g_jj'}) gives 
\begin{eqnarray}
\nonumber  
\tilde{g}_{jj'}(i\omega_n)\approx - 
\frac{m_*}{6\pi^2\vert {\bf r}_j-{\bf r}_{j'}\vert^2} \left\{\frac{1}{2} \ln\left[ 
\left(1-\frac{\pi^2}{2 m_* \vert {\bf r}_j-{\bf r}_{j'}\vert^2 \epsilon_F^2}\right)^2 
\right.\right. && \\
\nonumber 
\left. \left.  
+\frac{\pi^4\omega_n^2}{4 m_*^2 \vert {\bf r}_j-{\bf r}_{j'}\vert^2 \epsilon_F^4} \right] +i\arctan\left(\frac{\pi^2\omega_n}{2 m_* \vert {\bf r}_j-{\bf r}_{j'}\vert^2 \epsilon_F^2-\pi^2}\right)\right\}. && \\
\label{g-eval}
\end{eqnarray}
Here $m_*$ denotes the effective mass of an electron, and $\epsilon_F$ denotes the Fermi energy. 
Perturbative evaluation of RKKY interaction in terms of parameters $g_{jj'}$ results in the expression 
\begin{equation}
H_{\mathrm{RKKY}}=\frac{J^2}{2} \left[T\sum_{\omega_n} \tilde{g}_{jj'}(i\omega_n) \tilde{g}_{j'j}(i\omega_n)\right] 
{\bf S}_j\cdot {\bf S}_{j'}.
\label{RKKY}
\end{equation}
One can see from (\ref{g-eval}) that $\tilde{g}_{jj'}$ takes complex values, and the relation between its real and imaginary parts depends on the distance between the atoms 
$\vert {\bf r}_j-{\bf r}_{j'}\vert$ and on the value of $\omega_n$. The dependence on the interatomic distance leads to the changing sign of the RKKY interaction. In particular, it follows from (\ref{RKKY}) that real-valued $\tilde{g}_{jj'}$ corresponds to the antiferromagnetic RKKY coupling. It is convenient to parameterize $\tilde{g}_{jj'}$ as 
$\tilde{g}_{ij}(i\omega_n)=-i {\mathrm{sgn}}(\omega_n) g_{jj'}( \omega_n)$. 
In what follows we consider a cluster in the form of an isosceles triangle with AFM RKKY coupling by choosing the explicit form of the matrix $g_{jj'}$ as follows 
\begin{equation}
{\bf g}=\left( \begin{array}{ccc}
g_0 & i g_1 & i g_2 \\
i g_1 & g_0 & i g_1 \\
i g_2 & i g_1 & g_0 
\end{array}
\right). 
\label{matrix_g}
\end{equation}
Minimizing the free energy (\ref{F}), we 
obtain mean field equations for the fields $b_j$ that for low temperatures can be written as  
\begin{equation}
b_j=\frac{J}{6\pi}\sum_{i=1}^3 \left\{\ln\left(\frac{\Lambda_0}{\vert \lambda_i\vert}\right)\frac{\partial\lambda'_i}{\partial b_j}+ \arctan\left(\frac{\lambda''_i}{\lambda'_i}\right) \frac{\partial\lambda''_i}{\partial b_j}\right\}. 
\label{meanfield}
\end{equation}
Here $\Lambda_0$ is a high-energy cutoff of the order of the band-width, which is  typical for the Kondo problem,  and $\lambda_i=\lambda'_i+i \lambda''_i$ are the eigenvalues of the matrix ${\bf BgB}$. The mean field solutions can further be used to determine the local one particle density of states in presence of magnetic impurities. The single particle Green's function in the mean field approximation can be written as 
\begin{equation}
{\bf G}(\epsilon; {\bf k}, {\bf k'})={\bf G}^0(\epsilon; {\bf k})\delta_{{\bf kk'}} +{\bf G}^0 (\epsilon; {\bf k})
{\bf T}(\epsilon) {\bf G}^0 (\epsilon; {\bf k'}). 
\label{G-meanfield}
\end{equation}
Here the matrix elements of the transfer matrix ${\bf T}(\epsilon)$ can be calculated as 
\begin{equation}
T_{ij}(\epsilon)=\sum_{n=1}^3 b_i({\bf U}^{-1})_{in}(\epsilon+i\lambda_n)^{-1} {\bf U}_{nj} b_j,
\label{Tij}
\end{equation}  
where ${\bf U}$ is the matrix diagonalizing ${\bf BgB}$. The density of states at a point ${\bf r}$ is given as $\rho(\epsilon, {\bf r})=-\frac{1}{\pi} {\mathrm{Im}} G^R(\epsilon, {\bf r},  {\bf r})$. The expression for the additional density of states induced by magnetic impurities acquires especially simple form at the points where the magnetic atoms are situated. It reads 
\begin{equation}
\Delta \rho_i(\epsilon)=-\sum_{m,l,n}  
\frac{g_{im}b_m({\bf U}^{-1})_{ml} \lambda'_l {\bf U}_{ln} b_n g_{ni}}{(\epsilon-\lambda''_l)^2+(\lambda'_l)^2} . 
\label{deltarho}
\end{equation}  
The induced density of states has been calculated for a magnetic clusters  of different form by solving numerically the mean field equations (\ref{meanfield}). Taking into account the symmetry of the isosceles triangle, we assumed equal mean field  values of the field $b_1$ and $b_3$. 

In the case of a strongly asymmetric  isosceles triangle there is a clear tendency to formation of magnetic dimer with the increase of the asymmetry of the triangle. The results for the induced DoS at one of the atoms forming a dimer are shown in Fig. 
\ref{fig-dimer}. The peak in the density of states is suppressed with growing antiferromagnetic RKKY interaction. 
A much more complicated behavior is found by the cluster in form of an almost equilateral triangle. The antiferromagnetic RKKY interaction introduces frustration of the magnetic state of the cluster, which leads to a degenerate ground state. This degeneracy is very likely to be broken by a small asymmetry of the triangle. Such an asymmetry has been introduced for calculations shown  in Fig. \ref{fig-trimer}. The calculations are performed for an isosceles triangle with the form very close to the equilateral one.  We found that even weak symmetry breaking in presence of frustration leads to large changes in local  DoS at different atoms. The trimer again demonstrates a tendency to split into a dimer and a single atom. The density of states however bears signatures of strong AFM interaction between the atoms. 
One can distinguish two regimes, the Kondo regime at weak RKKY interaction and the regime of frustrated antiferromagnetic cluster at strong interatomic RKKY exchange. 
In the Kondo regime, there is a Kondo peak at the position of each atom in the trimer. The Kondo peak is centered at the Fermi enegy ($\epsilon=0$). 
   We found that weak correlations between magnetic moments that are mediated by RKKY interactions lead to the {\em enhancement} of the  Kondo peak (not shown).  

At strong RKKY interaction the regime of antiferromagnetic cluster is realized.    
DoS reflects the  separation of  the equilateral triangle into a dimer and a singled out atom. This is indicated by a strong drop in DoS on dimerized atoms accompanied by a splitting of the Kondo peak (see the curve for $g_1/g_0=0.6$ in the panel a) while there is a pronounced Kondo peak at the position of the separated atom (panel b). The splitting of the Kondo peak has also been observed in numerical simulations \cite{Savkin}. With further growth of RKKY interaction, the splitting of the Kondo peak at a dimer atom becomes asymmetric in energy, the peak in DoS shifts from the position of the Fermi level (the curve $g_1/g_0=0.6$ in the panel a). DoS on the singled out atom is shown in the panel b) to Fig. \ref{fig-trimer}. The Kondo peak  at the  singled out atom becomes broader and lower with growing RKKY coupling.  We attribute this peculiar behavior to the strong frustration in the equilateral triangular cluster with AFM nearest neighbor interaction. The shift of the Kondo peak from the position of the Fermi level ($\epsilon=0$) may indicate an appearance of a new phase for the triangular cluster which might demonstrate a non-Fermi-liquid behavior \cite{Affleck}. 

In conclusion, we proposed a theoretical treatment that takes into account Kondo correlations and RKKY interaction between localized  magnetic moments on equal footing. We applied this approach to triangular magnetic clusters and investigated the dependence of the local density of states on the geometry of the cluster. We found two qualitatively different regimes. In the regime of weak antiferromagnetic RKKY interaction, the RKKY coupling enhances Kondo peak in DoS at positions of magnetic atoms. In the opposite regime of strong RKKY interaction, the Kondo effect is strongly suppressed, and DoS is essentially featureless. 

We believe that our approach can further be applied to the magnetic clusters in the mixed valence regime and to the clusters with larger number of atoms.  The form of the mean field equations (\ref{meanfield}) and of the local density of states 
(\ref{deltarho}) remains unchanged, while the matrix $\mathbf{g}$ (\ref{matrix_g}) for a cluster of $N$ atoms becomes $N$-dimensional. Diagonalizing the matrix $\mathbf{B}\mathbf{g}\mathbf{B}$ thus becomes more difficult but should be numerically accomplished without severe problems.   

\begin{acknowledgments}  
Authors acknowledge Financial support from DFG through Sonderforschungsbereich 668. 
\end{acknowledgments}

\end{document}